\documentclass[letterpaper]{article} 
\usepackage{aaai23}  
\usepackage{times}  
\usepackage{helvet}  
\usepackage{courier}  
\usepackage[hyphens]{url}  
\usepackage{graphicx} 
\usepackage{natbib}  
\usepackage{caption} 
\usepackage{algorithm}
\usepackage{algorithmic}
\usepackage[most]{tcolorbox}
\usepackage{newfloat}
\usepackage{listings}
\DeclareCaptionStyle{ruled}{labelfont=normalfont,labelsep=colon,strut=off} 
\floatstyle{ruled}
\newfloat{listing}{tb}{lst}{}
\floatname{listing}{Listing}
\usepackage{microtype}
\usepackage{graphicx}
\usepackage{subfigure}
\usepackage{booktabs} 
\usepackage{multirow}

\usepackage{amsmath}
\usepackage{amssymb}
\usepackage{mathtools}
\usepackage{amsthm}

\usepackage{xpatch}
\makeatletter

\xpatchcmd{\algorithmic}
  {\newcommand{\STATE}{\ALC@it}}
  {\newcommand{\STATE}{\@ifstar\STATEstar\STATEnostar}}
  {}{}
\newcommand{\STATEstar}{\item[]}
\newcommand{\STATEnostar}{\ALC@it}

\theoremstyle{plain}
\newtheorem{theorem}{Theorem}

\newtheorem{definition}[theorem]{Definition}

\theoremstyle{remark}

\newtheorem{example}{Example}

\title{On-Device Model Fine-Tuning with Label Correction in Recommender Systems}

\author {
    Yucheng Ding\textsuperscript{\rm 1},
    Chaoyue Niu\textsuperscript{\rm 1},
    Fan Wu\textsuperscript{\rm 1},
    Shaojie Tang\textsuperscript{\rm 2},
    Chengfei Lyu\textsuperscript{\rm 3}, and
    Guihai Chen \textsuperscript{\rm 1}
}
\affiliations {
    \textsuperscript{\rm 1} Shanghai Jiao Tong University, China\\
    \textsuperscript{\rm 2} University of Texas at Dallas, USA\\
    \textsuperscript{\rm 3} Alibaba Group, China\\
    \{yc.ding, rvince, wu-fan, gchen\}@sjtu.edu.cn,
    \{shaojie.tang\}@utdallas.edu,
    \{chengfei.lcf\}@alibaba-inc.com
}

\usepackage{bibentry}

\begin{document}

\maketitle

\begin{abstract}
To meet the practical requirements of low latency, low cost, and good privacy in online intelligent services, more and more deep learning models are offloaded from the cloud to mobile devices. To further deal with cross-device data heterogeneity, the offloaded models normally need to be fine-tuned with each individual user's local samples before being put into real-time inference. In this work, we focus on the fundamental click-through rate (CTR) prediction task in recommender systems and study how to effectively and efficiently perform on-device fine-tuning. We first identify the bottleneck issue that each individual user's local CTR (i.e., the ratio of positive samples in the local dataset for fine-tuning) tends to deviate from the global CTR (i.e., the ratio of positive samples in all the users' mixed datasets on the cloud for training out the initial model). We further demonstrate that such a CTR drift problem makes on-device fine-tuning even harmful to item ranking. We thus propose a novel label correction method, which requires each user only to change the labels of the local samples ahead of on-device fine-tuning and can well align the locally prior CTR with the global CTR. The offline evaluation results over three datasets and five CTR prediction models as well as the online A/B testing results in Mobile Taobao demonstrate the necessity of label correction in on-device fine-tuning and also reveal the improvement over cloud-based learning without fine-tuning.
\end{abstract}

\section{Introduction}

In recent years, deep learning has been widely deployed in industrial recommender systems. In addition, due to the stringent latency requirement on returning recommendations upon receiving each request from millions or even billions of user (e.g., in hundreds of milliseconds), more and more recommendation models are first trained on the cloud (e.g., with standard learning algorithms or meta learning algorithms) and then offloaded to mobile devices for real-time inference. Such an on-device learning paradigm leverages the natural advantages of mobile devices being close to users and at data sources, thereby reducing latency and communication overhead, mitigating the cloud-side load, and keeping raw data with sensitive contents (e.g., user behaviors) on local devices.       




However, the ubiquitous issue of cross-device data heterogeneity in recommender systems makes the cloud-based global model non-optimal to directly serve each individual user. In particular, different users normally have diverse behavior patterns (e.g., daily active vs. monthly active, and different sequences of browsed, clicked, and purchased goods) and differentiated preferences (e.g., some users like sports-related goods, while some users prefer foods). This implies that each individual user's local data distribution tend to deviate from the global data distribution (i.e., a mixture of all the users' data distributions). As a result, the cloud-base model, which is optimized over the global data distribution, may not perform very well for all the users.


To deal with cross-device data heterogeneity, fine-tuning the cloud-based model on each mobile device with the corresponding user's local samples is a potential solution, considering both effectiveness and efficiency. On the one hand, on-device fine-tuning can adapt to each user's data distribution and achieve extremely model personalization, one personalized recommendation model for one user. On the other hand, given the fact that the size of each individual user's local samples is small, while the cloud-based model is nearly optimal, fine-tuning tends to require only a few model iterations, the overhead of which is affordable to resource-constraint mobile devices. 

\begin{figure}[!t]
\centering
\includegraphics[width=0.9\columnwidth]{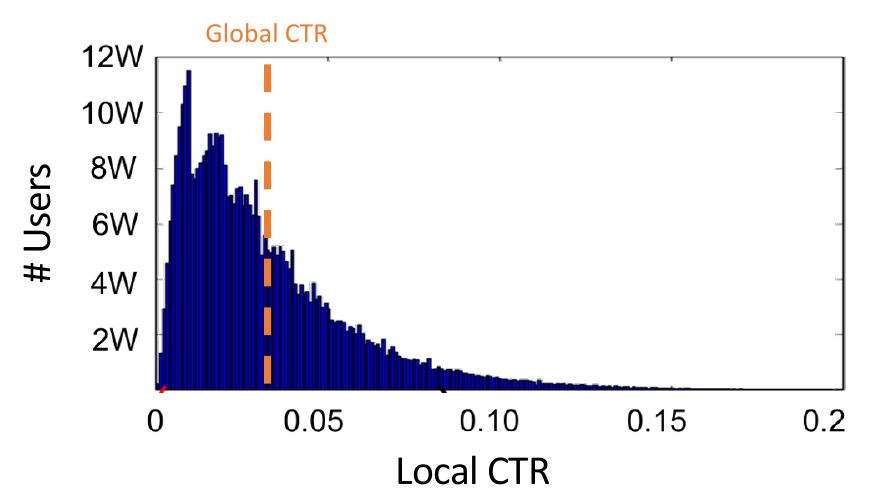}
\caption{The distribution of Mobile Taobao users' local CTRs (some tail local CTRs are not displayed), which is based on the statistics over 5 million Taobao users’ 3-day data on the homepage recommendation.}
\label{ctr_stats}
\end{figure}

In this work, we focus on the basic click through rate (CTR) prediction task in recommender systems, which predicts whether a user will click an candidate item or not given the user's profile and the user's historic behavior sequence, and identify atypical issues in on-device fine-tuning. In particular, when each individual user fine-tunes a CTR prediction model with its local samples, the model update is sparse rather than dense (e.g., in the context of most computer vision applications), which was also called  ``submodel update'' in ~\cite{niu_mobicom20}. The major reason is that a certain user normally interacts only with a small number of items, and his/her local samples involve a small subspace of the full feature space and will update a small part of the full CTR prediction model (e.g., the embedding vectors of a few local items) in the phase of fine-tuning. In addition to sparse model update, we also observe a brand new issue, called ``CTR drift'', which means that each individual user's local CTR (i.e., the ratio of positive samples in the local dataset for fine-tuning) may deviate from the global CTR (i.e., the ratio of positive samples in the global dataset on the cloud for training out the initial model). Specifically, we collect 3-day data of 5 million users from Mobile Taobao, which is the largest online shopping platform in China, and depict in Figure \ref{ctr_stats} the statistics of the local CTRs and the global CTR. We can observe from Figure \ref{ctr_stats} that the local CTRs of most users deviate from the global CTR.


We further demonstrate that the CTR drift will deteriorate rather than benefit the performance of the CTR prediction model after local sparse fine-tuning. In particular, for a user with his/her local CTR being higher (resp., lower) than the global CTR, the sparsely updated parameters of the fine-tuned model tend to be larger (resp., smaller) than the other parameters, which causes some items having too high (resp., low) predicted CTRs and thus disrupts the ranking result of the candidate items. To deal with this problem, we propose a novel label correction method for on-device fine-tuning (``LCFT'' for short), which is effective, efficient, and yet simple in implementation. In detail, LCFT just needs to correct the labels of local samples, such that the local equivalent CTR, which is defined as the optimal prior CTR that minimizes the training loss over each individual user's dataset, will be equal to the global CTR. Intuitively, the local equivalent CTR determines the magnitude of the updated parameters. Therefore, the label correction procedure can ensure that the updated parameters and the other non-updated parameters will be in the equal magnitude. Based on the alignment of the equivalent local CTR and the global CTR, we provide the theoretical choice of the corrected labels for each individual user. In addition to the way of soft setting, the choice of label correction can also be set to hyperparameters for hard searching in practice.


We summarize the contributions of this work as follows:
\begin{itemize}
    \item We study how to effectively and efficiently perform on-device fine-tuning for the CTR prediction model in recommender systems, and identify the bottleneck issue of CTR drift for the first time.
    \item To mitigate the negative effect of CTR drift on the local sparse fine-tuning, we propose a novel label correction method LCFT, which requires each user only to change the labels of local samples before fine-tuning. Theoretically, LCFT aligns the local equivalent CTR with the global CTR and further keeps the updated parameters and the other non-updated parameters in the equal magnitude. 
    \item We extensively evaluate LCFT on five recommendation models over the public MovieLens 20M and Amazon Electronics datasets, as well as an industrial 7-day dataset collected from Mobile Taobao. The offline evaluation results demonstrate that LCFT outperforms cloud-based learning and purely on-device fine-tuning, validating the necessity of label correction.
    \item We deploy LCFT in the homepage recommendation of Mobile Taobao for online A/B testing. The online results reveal that LCFT can indeed improve the personalized recommendation performance in terms of both CTR and user activenss. Specifically, LCFT improves CTR by 1.79\% compared with cloud-based learning. 
\end{itemize}

\section{Related Work}




\subsection{On-Device Inference and Fine-Tuning} 

With the rapid development of mobile devices and the advance of model compression algorithms, how to enable on-device learning emerges as a hot and promising topic. 

Much previous work has devoted to on-device inference, where models are first trained on the cloud and then offloaded the mobile devices for inference. \citet{deepeye} and \citet{deepmon} designed DeepEye and  DeepMon to support the on-device inference of computer vision models. Siri, the voice assistant developed by Apple, adopted on-device inference to improve the text-to-speech synthesis process \cite{siri1}. \citet{lv_walle} built an end-to-end, general-purpose, and large-scale production system Walle, having deployed many recommendation, computer vision, and natural language tasks to mobile APPs.

Besides on-device inference, on-device fine-tuning is also strongly motivated to tackle data heterogeneity among users. In particular, it is normally non-optimal to directly use the cloud-based global model, trained by standard algorithms or meta-learning algorithms~\cite{maml, person_maml}, to serve each individual user in real time. The technique of fine-tuning, which was initially proposed in the context of big model pre-training to quickly adapt to a specific downstream task~\cite{bert, mae}, is quite convenient and efficient to be applied to resource-constraint mobile devices. For example, \citet{deeptype} focused on the next-word prediction task and proposed to leverage on-device fine-tuning for model personalization, thereby improving prediction accuracy. Their focus was on how to reduce overhead by vocabulary and model compression. In contrast, we focus on the CTR prediction task in recommender systems and consider how to address the newly identified CTR drift for unbiased optimization.



\subsection{Cross-Device Learning}

In addition to on-device learning, where the learning tasks on different mobile devices are independent from each other, another line of existing work focused on cross-device learning, where multiple mobile devices perform joint optimization. Federated learning~\cite{McMahan_aistats17, li_mlsys20,cho_corr20, Li_iclr20, Karimireddy_icml20} is the most popular cross-device learning framework, which allows many users with mobile devices to collaboratively train a global model under the coordination of a cloud server without sharing their local data, such that data privacy can be well protected. Some other work studied different collaboration mechanisms between the cloud and multiple mobile devices to achieve model personalization. \citet{yan_kdd22} proposed MPDA, which requires a certain user retrieve some similar data from other users. From the perspective of domain adaption, other users' data function as large-scale source domains and are retrieved to augment the user’s local data as the target domain. \citet{gu_alibaba21} proposed CoDA, which lets each user train an on-device, two-class classifier for data augmentation. 

Complementary to federated learning, which trains out a global model without sharing raw data, on-device fine-tuning can be further applied for model personalization. In contrast to data sharing based work (e.g., \citet{yan_kdd22, gu_alibaba21}), on-device fine-tuning allows each individual user to use only his/her data on local devices, which preserves data privacy and is easy to be deployed in practice.



\subsection{Personalized Recommender Systems} 

The basic task of recommender systems is to rank different candidate items according to the metric of CTR. Before deep learning was introduced in recommender systems, logistic regression (LR) was the widely used shallow model. Later, \cite{net_wd} proposed Wide\&Deep, which combines the memorization strength of LR and the generalization ability of deep neural networks. DeepFM~\cite{net_fm} replaced the wide part with a factorization machine. PNN~\cite{net_pnn1,net_pnn2} introduced a production layer before the multi-layer perceptron. Deep interest network (DIN)~\cite{net_din} added an attention layer to extract the relevance of a user’s historical behaviors to a candidate item. For on-device recommendation,\cite{net_edgerec} designed EdgeRec, which introduced several real-time user behavior context, such as duration of item exposure, scroll speed and scroll duration for item exposure, and deleting reason. EdgeRec has achieved remarkable performance improvement and has been widely deployed in Mobile Taobao. 

The key focus of previous work in recommender systems was to design diverse model structures to represent user data better or to introduce more personalized features from users. However, it is difficult for a single global recommendation model to be optimal on each individual user's local data, making model personalization become a new trend. This work exactly studies in the way of on-device fine-tuning.

\section{Problem Formulation}

There are $u$ users with mobile devices in total, denoted as $\{1,2,\cdots,u\}$. The dataset of user $i$ for local fine-tuning is denoted as $D_i$. Considering data heterogeneity in practice, different users' datasets tend to follow different distributions. In this work, we focus on the CTR prediction task in recommender systems, which is a standard two-class classification problem. For user $i$'s dataset $D_i$, we let $n_i^+$ denote the size of positive samples, let $n_i^-$ denote the size of negative samples, and let $w_i \overset{\triangle}{=} \frac{n_i^+}{n_i^+ + n_i^-}$ denote the ratio of positive samples or the local CTR. The optimization objective of user $i$ is
\begin{align}\label{local_opt}
    L_i = \min_{{\bf h}_i} \sum_{(x,y)\in D_i} l\left({\bf h}_i\left(x\right),y\right),
\end{align}
where ${\bf h}_i$ denotes the local recommendation model, $x$ and $y$ denote the input features and the label of a training sample, and $l(\cdot,\cdot)$ denotes loss function, typically mean square error or cross-entropy loss.

\subsection{Cloud-Based Learning for Model Initialization}

The initial model of each user's local fine-tuning normally comes from cloud-based training, which optimizes a global model over all the users' datasets with the objective:
\begin{align}\label{cloud_opt}
    L = \min_{{\bf h}}\frac{1}{u} \sum_{i=1}^u \sum_{(x,y)\in D_i} l\left({\bf h}\left(x\right),y\right).
\end{align}
Equation \ref{cloud_opt} reveals that the discrepancy among different $D_i$'s (i.e., cross-device data heterogeneity) limits the performance of the cloud-based global model ${\bf h}$, when serving each user. Formally, the globally optimal model ${\bf h}$ may be non-optimal for each user's dataset $D_i$, and $L\le \frac{1}{u}\sum_{i=1}^u L_i$. 

\subsection{On-Device Fine-Tuning with CTR Drift Problem}

To break the dilemma of cloud-based learning, on-device fine-tuning is a potential solution. In particular, each user $i$ will continue to optimize the global model ${\bf h}$ with its local dataset $D_i$ by minimizing equation \ref{local_opt} independently. In the context of recommender systems, we observe two atypical phenomenons in practice: (1) {\bf Sparse Model Update}: each user fine-tunes the deep recommendation model with its local dataset, updating only part of model parameters rather that the full model. Specifically, for the embedding layer in the CTR prediction model, only the embedding vectors corresponding to the items in each user $i$'s local dataset $D_i$ will be updated; and for the other layers (e.g., MLP with ReLu activation function), only part of the neurons with non-zero inputs will be updated; and (2) {\bf CTR Drift}: each individual user $i$'s local CTR $w_i = \frac{n_i^+}{n_i^+ + n_i^-}$ deviates from the global CTR $w_g \overset{\triangle}{=} \frac{\sum_{i=1}^{u} n_i^+}{\sum_{i=1}^{u} \left(n_i^+ + n_i^-\right)}$, or the ratio of positive samples in the local dataset $D_i$ for on-device fine-tuning differs from the ratio of positive samples in the global dataset $\bigcup_{i=1}^{u} D_i$ for training out the initial model on the cloud. Specifically, by analyzing over 5 million users's data collected from Mobile Taobao in the period of 3 days, we find that the users's local CTRs follow a long-tail distribution as shown in Figure \ref{ctr_stats}, and the local CTRs of most users are far from the global CTR. Further, we find that the issue of CTR drift will seriously degrade the performance of each user's sparse fine-tuning. We give the detailed illustrations as follows.




From the perspective of a certain user, if its local CTR is higher (resp., lower) than the global CTR, the updated model parameters tend to be larger (resp., smaller) than those parameters that are not updated. The discrepancy between the updated and the non-updated parameters will lead to bad prediction performance, because the fine-tuned model tends to output higher (resp., lower) CTR predictions for the items involving more updated parameters. As a result, even if the fine-tuned model has a higher test accuracy and a lower test loss, the ranking result of different items for generating final recommendations may become worse. We take an example for more intuitive illustration. 

\begin{example}\label{ctrdrift_ex}
The global CTR is 0.5, and user $i$'s local CTR is 0.25. For user $i$, suppose that the true CTRs of three items $I_1$, $I_2$, and $I_3$ are $0.3$, $0.1$, and $0.2$, respectively, and the CTRs of the three items predicted by the cloud-based model are $0.5$, $0.35$, and $0.6$, respectively. Both $I_1$ and $I_3$ involve $m$ samples in the local dataset $D_i$, and are clicked by $0.3m$ and $0.2m$ times, respectively, while $I_2$ does not appear in $D_i$. Assume that by on-device fine-tuning, the local model can precisely predict the CTRs of $I_1$ and $I_3$ while none of the model parameters related to $I_2$ is updated. Then, the CTRs of $I_1$, $I_2$, and $I_3$ predicted by the fine-tuned model become $0.3$, $0.35$, and $0.2$, respectively. 
\end{example}

In Example \ref{ctrdrift_ex}, the ranking result of the cloud-based model is $\{I_3, I_1, I_2\}$, whereas the ranking result of the fine-tuned model is $\{I_2, I_1, I_3\}$. Compared with the true ranking result $\{I_1, I_3, I_2\}$, the fine-tuned model correctly ranks $I_1$ and $I_3$, but incorrectly ranks $I_2$ in the first place. We note that the ranking result after on-device fine-tuning even becomes worse, although the fine-tuned model has a lower test loss than the cloud-based model.

\section{Algorithm Design and Implementation}

To mitigate the effect of CTR drift on the local fine-tuning, we propose a novel label correction method, which is simple in implementation and quite effective and efficient. The key idea is to correct the labels of local samples such that the equivalent CTR with respect to the local optimization objective, formally defined in Definition \ref{eq_ctr}, is consistent of the global CTR after label correction. Therefore, the discrepancy between the updated and the non-updated model parameters after on-device fine-tuning can also be mitigated. 

\begin{definition}\label{eq_ctr}
User $i$'s equivalent CTR ${y^0_i}^*$ is the prior CTR that minimizes the loss over its training dataset, formally,
\begin{align}
{y^0_i}^* = \arg\min_{y^0_i} \sum_{(x,y) \in D_i} l(y^0_i, y),
\end{align}
where $L_i^0\overset{\triangle}{=} \sum_{(x,y) \in D_i} l(y^0_i, y)$ denotes the training loss by using the prior CTR $y_i^0$ as the predicted labels.
\end{definition}

By Definition \ref{eq_ctr}, we next calculate the equivalent CTR ${y^0_i}^*$ for user $i$. We consider that $l(\cdot,\cdot)$ takes the mean square error or the cross-entropy loss function. We let $\alpha_i$ and $\beta_i$ ($\alpha_i>\beta_i$) denote the corrected labels of a positive sample and a negative sample, respectively. First, for the mean square error, the training loss is  
\begin{equation}
\begin{aligned}
    L_i^0 =& n_i^+ \left(\alpha_i - y_i^0 \right)^2 + n_i^- \left(\beta_i-y_i^0\right)^2.
\end{aligned}
\end{equation}
To minimize $L_i^0$, we let $\frac{\partial L_i^0}{\partial y_i^0} = 0$ and have
\begin{align}
    -2n_i^+(\alpha_i-y_i^0) - 2n_i^-(\beta_i-y_i^0) = 0.
\end{align}
Given the local CTR $w_i = \frac{n_i^+}{n_i^+ + n_i^-}$, we have the equivalent CTR ${y_i^0}^*$ in the formula of the local CTR $w_i$ and the corrected labels $\alpha_i$ and $\beta_i$:
\begin{align}\label{mse_label}
    {y_i^0}^* = w_i \alpha_i + (1-w_i) \beta_i.
\end{align}
Similarly, for the cross-entropy loss, the training loss is
\begin{equation}
\begin{aligned}
    L_i^0 = &n_i^+ \left(\alpha_i \ln{y_i^0} + (1-\alpha_i)\ln{(1-y_i^0)} \right) \\
            &+ n_i^- \left(\beta_i \ln{y_i^0} + (1-\beta_i)\ln{(1-y_i^0)} \right).
\end{aligned}
\end{equation}
We minimize $L_i^0$ by letting $\frac{\partial L_i^0}{\partial y_i^0} = 0$ and have
\begin{align}
    n_i^+\left( \frac{\alpha_i}{y_i^0} - \frac{1-\alpha_i}{1-y_i^0} \right) 
    + n_i^- \left( \frac{\beta_i}{y_i^0} - \frac{1-\beta_i}{1-y_i^0} \right) = 0. 
\end{align}
Given $w_i = \frac{n_i^+}{n_i^+ + n_i^-}$, we have the equivalent CTR
\begin{align}\label{ce_label}
    {y_i^0}^* = w_i \alpha_i + (1-w_i) \beta_i,
\end{align}
which is in the same format as for the mean square error. 


After obtaining user $i$'s equivalent CTR ${y_i^0}^*$ in equation \ref{mse_label} or \ref{ce_label}, we let it to be equal to the global CTR $w_g$:
\begin{align}\label{CTR_l=g}
    w_i \alpha_i + (1 - w_i) \beta_i = w_g.
\end{align}
We then can obtain the corrected labels $\alpha_i$ and $\beta_i$ using the global CTR $w_g$ and the local CTR $w_i$. Since equation~\ref{CTR_l=g} is indeterminate, we just give two trivial solutions by keeping the label of negative samples unchanged $\beta_i = 0$ or keeping the label of positive samples unchanged $\alpha_i = 1$, and we have
\begin{align}\label{sol_label}
\left\{
\begin{aligned}
\alpha_i &= \frac{w_g}{w_i} \\
\beta_i &= 0
\end{aligned}
\right.
\ \ \ \text{or} \ \ \ 
\left\{
\begin{aligned}
\alpha_i &= 1 \\
\beta_i &= \frac{w_g - w_i}{1-w_i}
\end{aligned}
\right.
.
\end{align}

To intuitively demonstrate the effect of label correction, we still examine Example \ref{ctrdrift_ex} using the first choice of corrected labels\footnote{For the second choice of label correction, the local predicted CTRs of $I_1$ and $I_3$ become 0.53 and 0.47, and the ranking result is also correct.} $\alpha_i = \frac{w_g}{w_i}$ and $\beta_i = 0$. Given $w_i = 0.25$ and $w_g=0.5$ in Example \ref{ctrdrift_ex}, user $i$ will correct the positive label to $\alpha_i = 2$ and keep the negative label unchanged $\beta_i = 0$, which implies that the labels of $I_1$-related and $I_3$-related clicked samples become 2, while the labels of non-clicked samples keep 0. By on-device fine-tuning, the local model learns the optimal CTRs of $I_1$ and $I_3$, which are 0.6 and 0.4, respectively, and the predicted CTR of $I_2$ will remain 0.35. After label correction, the fine-tuned model can correctly rank the candidate items as $\{I_1, I_3, I_2\}$.

\begin{algorithm}[!t]
	\caption{Label Correction Based Fine-Tuning (LCFT)}\label{lcft}
	\begin{algorithmic}[1]
	    \REQUIRE The cloud-based model ${\bf h}$, the global CTR $w_g$;
	    \FOR{each client $i\in \{1,2,\cdots, u\}$ in parallel}
	        \STATE Collects training dataset $D_i$ generated on the device and obtains some statistics: the size of positive samples $n_i^+$, the size of negative samples $n_i^-$, and the local CTR $w_i \leftarrow \frac{n_i^+}{n_i^+ + n_i^-}$;
	        \STATE* {\tt /* Label correction */}
	        \STATE \colorbox{blue!30}{Gets $\alpha_i,\beta_i$ from the cloud;\ \textbf{(Hard Correction)}}
	        \IF{$w_i > w_g$}
	             \STATE \colorbox{red!30}{$\alpha_i \leftarrow \frac{w_g}{w_i}$, $\beta_i\leftarrow 0$;\  \textbf{(Soft Correction 1)}}
	             \STATE \colorbox{green!30} {$\alpha_i \leftarrow 1$, $\beta_i\leftarrow \frac{w_g - w_i}{1-w_i}$;\ \textbf{(Soft Correction 2)}}
            \ELSE
                 \STATE \colorbox{red!30} {$\alpha_i \leftarrow 1$, $\beta_i\leftarrow \frac{w_g - w_i}{1-w_i}$;\ \textbf{(Soft Correction 1)}}
	             \STATE \colorbox{green!30}{$\alpha_i \leftarrow \frac{w_g}{w_i}$, $\beta_i\leftarrow 0$;\  \textbf{(Soft Correction 2)}}
	        \ENDIF
	        \STATE Corrects the labels of positive and negative samples in $D_i$ to $\alpha_i$ and $\beta_i$, respectively;
	        \STATE* {\tt /* Fine-tuning */}
	        \STATE Initializes the local model ${\bf h}_i \leftarrow {\bf h}$;
	        \STATE Fine-tunes $w_i$ over $D_i$ with corrected labels;
	    \ENDFOR
	\end{algorithmic}
\end{algorithm}

We finally present the implementation details of the proposed label correction based fine-tuning method (LCFT)  in Algorithm \ref{lcft}. For each user $i$, it first collects the training data $D_i$ generated on the mobile device and obtain some statistical information (line 2). Based on the statistics and equation \ref{sol_label}, user $i$ can obtain the corrected labels and then perform label correction for $D_i$ (lines 3-11). Finally, each user $i$ can perform on-device fine-tuning, i.e., first initialize ${\bf h}_i$ with the cloud-based global model ${\bf h}$ and then fine-tunes ${\bf h}_i$ over $D_i$ with the corrected labels.

In practice, we propose three label correction strategies for online deployment, two soft correction strategies and one hard correction strategy, as highlighted in different colors. (1) For the two soft choices of corrected labels in equation \ref{sol_label}, we can also determine which specific choice based on the offline experiment before online deployment. Intuitively, soft correction 1 reduces the gap between the positive and negative labels, makes the loss function smoother, and is suitable for the case where the cloud-based model is close to the locally optimal model. In contrast, soft correction 2 amplifies the gap and is suitable for the case where the cloud-based model is far from the locally optimal model. (2) For the hard correction strategy, we can determine the corrected labels $\alpha_i,\beta_i$ based on the offline experiment as well as the local CTRs of each user during online fine-tuning. In particular, we can collect user logs and corresponding item features to build an offline dataset. Based on the offline dataset, we can group the users based on their local CTRs (i.e., different user groups have different CTR intervals), and grid-search the optimal choice of label correction for each group. Then, the cloud server can store the CTR intervals and the corresponding corrected labels. During online deployment, the mobile device can send its latest local CTR, and the cloud server will return the corresponding corrected labels. 

\section{Offline and Online Evaluations}

We extensively evaluate the proposed label correction design over both public and industrial datasets. We also deploy our design in the recommender system of Mobile Taobao and conduct online A/B testing.  

\subsection{Evaluation on Public Datasets}

We first evaluate over two public datasets. The statistics about users, samples, and the global CTRs are shown in Table \ref{data-stats}. The first dataset is {\bf MovieLens 20M}\footnote{https://grouplens.org/datasets/movielens/20m/}~\cite{movielens}, which contains 20,000,263 ratings from 138,493 users for 27,278 movies with 21 categories. The original user ratings of movies range from 0 to 5. We label the samples with the ratings of 4 and 5 to be positive and label the rest to be negative. Regarding the features of each user’s sample, we take the user’s ID, the historical sequence of positively rated movies, the historical sequence of negatively rated movies, as well as a candidate movie to be recommended and its tag genome. Each movie is represented by a unique ID and a category ID. The tag genome of a candidate movie, provided by the dataset, is a 1,128-dimensional vector and comprises the candidate movie’s relevance scores with respect to all tags. For the split of each user’s training and test sets, we take the samples with the timestamps no more than 1,225,642,324 into the training set and take the remaining samples into the test set. We evaluate on the 5,677 users who have both training and testing data. The second dataset is {\bf Amazon Electronics}\footnote{https://jmcauley.ucsd.edu/data/amazon/}, which contains 1,689,188 reviews contributed by 192,403 users for 63,001 products with 1,361 categories. For each user’s sample, the label is whether a candidate (i.e., reviewed or negatively sampled) product is reviewed or not, and the features include the user ID, the historical sequence of reviewed products, and the candidate product. Each product is represented by a product ID and a category ID. Regarding the split of the training and test sets, the samples with the timestamps no more than 1,385,078,400 fall into the training set, while the rest falls into the test set. We evaluate on the 180,342 users who have both training and testing data.

\begin{table}[!t]
	\caption{Statistics of two public datasets and one industrial dataset after pre-processing.}\label{data-stats}
	\begin{center}
		\resizebox{\columnwidth}{!}{
			\begin{tabular}{cccc}
				\toprule
				 & \#\ Users & \#\ Samples & Global CTR \\
				\cmidrule{2-4}
                MovieLens & 5,677 & 3,229,373 & 0.51 \\
                Amazon & 180,342 & 6,227,300  & 0.17 \\
                Mobile Taobao & 30,682 & 23,655,827 & 0.04 \\
				\bottomrule
			\end{tabular}
		}		
	\end{center}
\end{table}

\begin{table*}[!t]
    \caption{Offline user-level average AUCs of LCFT and the baselines.}
    \label{public_aucs}
    \centering
    \resizebox{1.95\columnwidth}{!}{
    \begin{tabular}{cccccccc}
        \toprule
        Dataset & Model & Cloud & Local & LCFT & Local vs. Cloud & LCFT vs. Cloud & LCFT vs. Local \\
        \cmidrule{1-8}
        \multirow{5}{*}{MovieLens} & LR & 0.624 & 0.624 & 0.624 & +0.00\% & +0.01\% & +0.01\% \\
         & Wide\&Deep & 0.652 & 0.657 & 0.660 & +0.77\% & +1.23\% & +0.46\% \\
         & DeepFM & 0.661 & 0.665 & 0.668 & +0.61\% & +1.06\% & +0.45\% \\
         & PNN & 0.671 & 0.675 & 0.676 & +0.60\% & +0.75\% & +0.15\% \\
         & DIN & 0.681 & 0.684 & 0.686 & +0.44\% & +0.73\% & +0.29\% \\
        \cmidrule{1-8}
        \multirow{5}{*}{Amazon} & LR & 0.678 & 0.679 & 0.679 & +0.04\% & +0.05\% & +0.00\% \\
         & Wide\&Deep & 0.764 & 0.772 & 0.778 & +1.05\% & +1.83\% & +0.78\% \\
         & DeepFM & 0.724 & 0.725 & 0.725 & +0.11\% & +0.15\% & +0.03\% \\
         & PNN & 0.755 & 0.754 & 0.755 & {\color{green}-0.07\%} & +0.07\% & +0.13\% \\
         & DIN & 0.791 & 0.791 & 0.794 & {\color{green}-0.11\%} & +0.38\% & +0.38\% \\ 
        \cmidrule{1-8}
        Taobao & EdgeRec & 0.614 & 0.614 & 0.617 & {\color{green}-0.05\%} & +0.49\% & +0.49\% \\
        \bottomrule
    \end{tabular}
    }
\end{table*}



We take five representative CTR prediction models, including {\bf LR}, {\bf Wide\&Deep}, {\bf DeepFM}, {\bf PNN}, and {\bf DIN}. We also introduce two baselines for comparison. One baseline is {\bf cloud-based learning} (abbreviated as ``{\bf Cloud}''), which trains the global model over all the users' data and is to verify the necessity of on-device fine-tuning; the other baseline is {\bf purely local fine-tuning} (abbreviated as ``{\bf Local}''), which directly lets each mobile device fine-tune the cloud-based model over the user's local dataset and is to verify the necessity of label correction in our design {\bf LCFT}.


Regarding the experimental settings, we choose mini-batch SGD as the optimization algorithm. For cloud-based learning over the MovieLens dataset, we set the batch size to 32 and train 2 epochs with the learning rate starting at 1 and decaying by 0.1 every epoch. For cloud-based learning over the Amazon dataset, we set the batch size to 1,024 and take the same settings for other hyperparameters. For on-device fine-tuning over two public datasets, we set the batch size to 32 and let each user train 1 epoch with the learning rate of 0.01 by default. Specially, for the Amazon dataset, we set the batch size to 16 in the fine-tuning of LR, and set the batch size to 4 and the learning rate to 0.001 in the fine-tuning of PNN. In addition, for offline evaluation, we adopt a golden metric in evaluating the performance of CTR prediction, called Area Under the ROC Curve (AUC). We use the user-level average AUC ~\cite{gauc1,gauc2}, defined as
\begin{align}
    AUC_{Avg} = \frac{\sum_{i=1}^u m_i AUC_i}{\sum_{i=1}^u m_i},
\end{align}
where $m_i$ denotes the size of user $i$'s test set, and $AUC_i$ is the AUC over user $i$'s individual testset.


\begin{figure}[!t]
\centering
\subfigure[Movielens]{
\includegraphics[width=0.472\columnwidth]{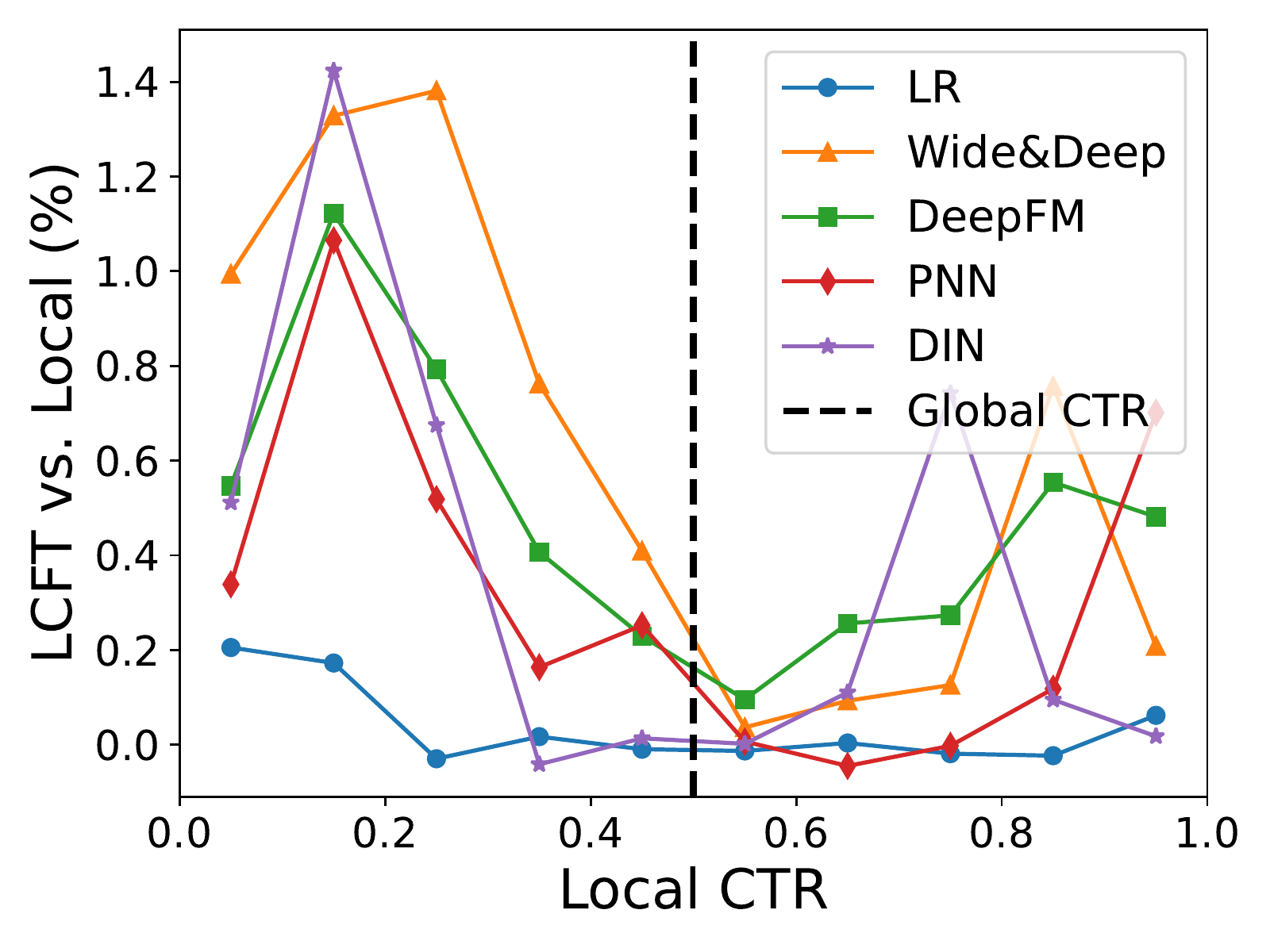}
}
\subfigure[Amazon]{
\includegraphics[width=0.472\columnwidth]{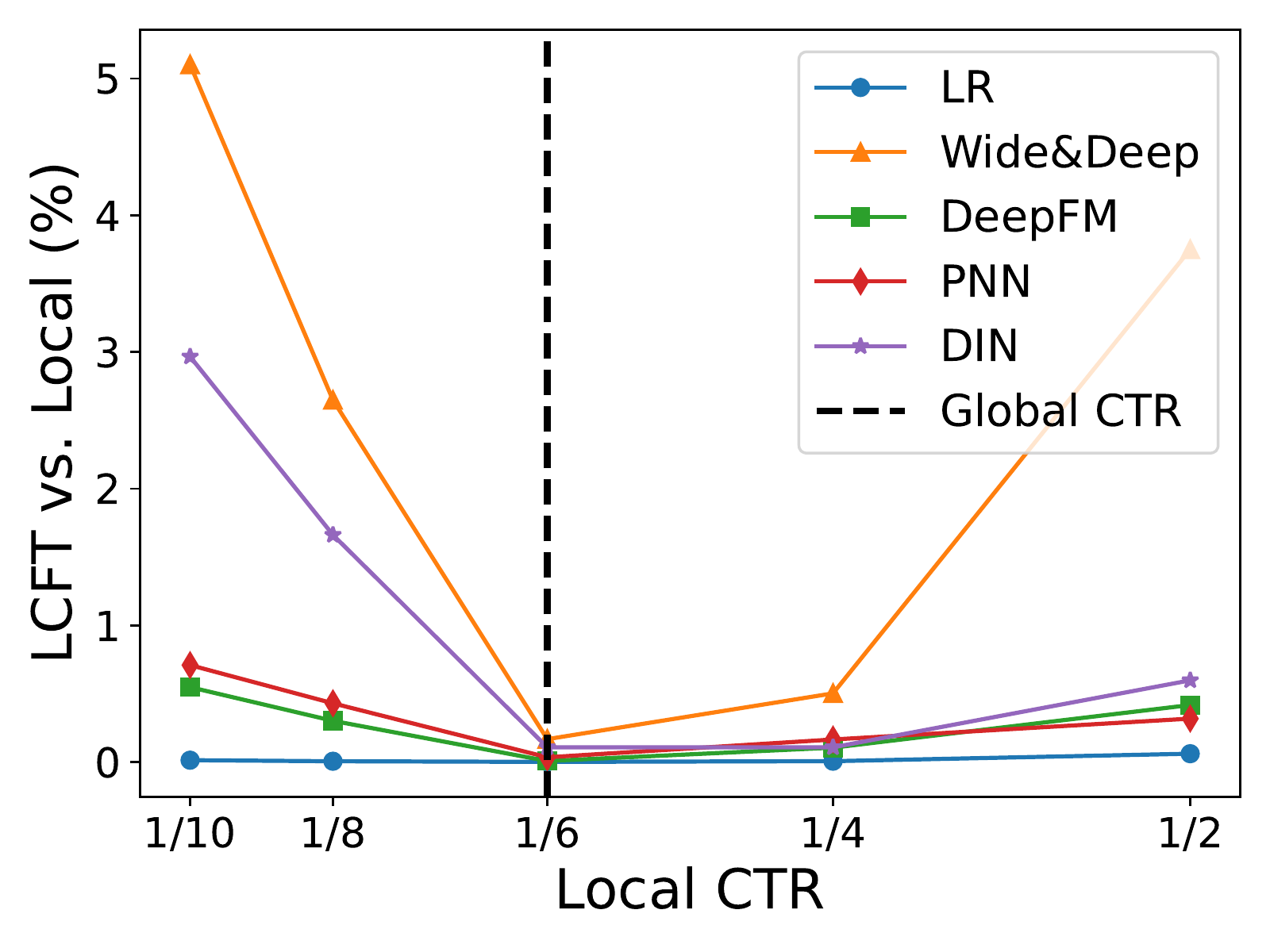}
}
\caption{The improvement of LCFT over purely local fine-tuning for different CTR drifts.}\label{datasets_gaps}
\end{figure}

\begin{figure}[!t]
\centering
\subfigure[Movielens]{
\includegraphics[width=0.472\columnwidth]{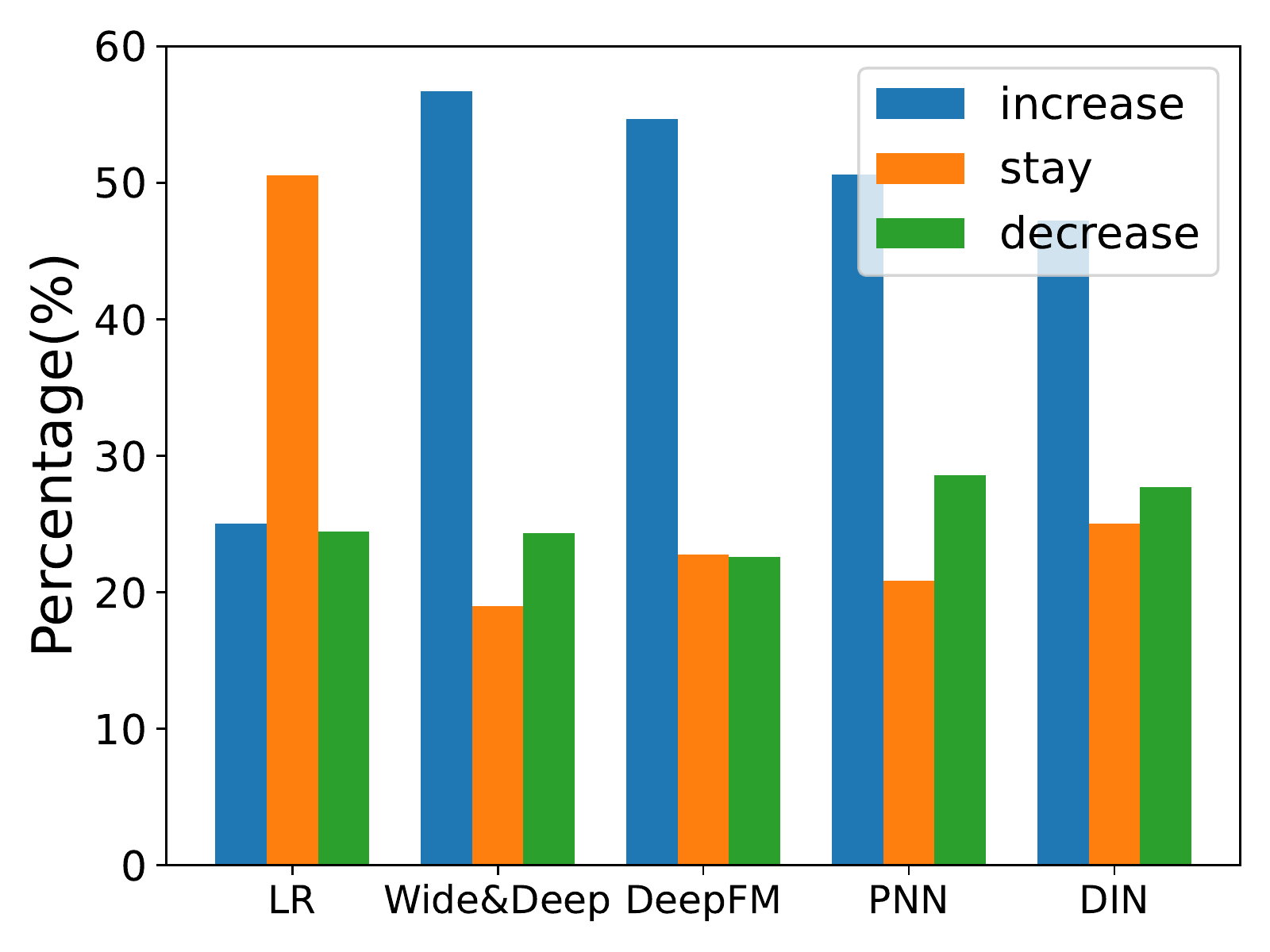}
}
\subfigure[Amazon]{
\includegraphics[width=0.472\columnwidth]{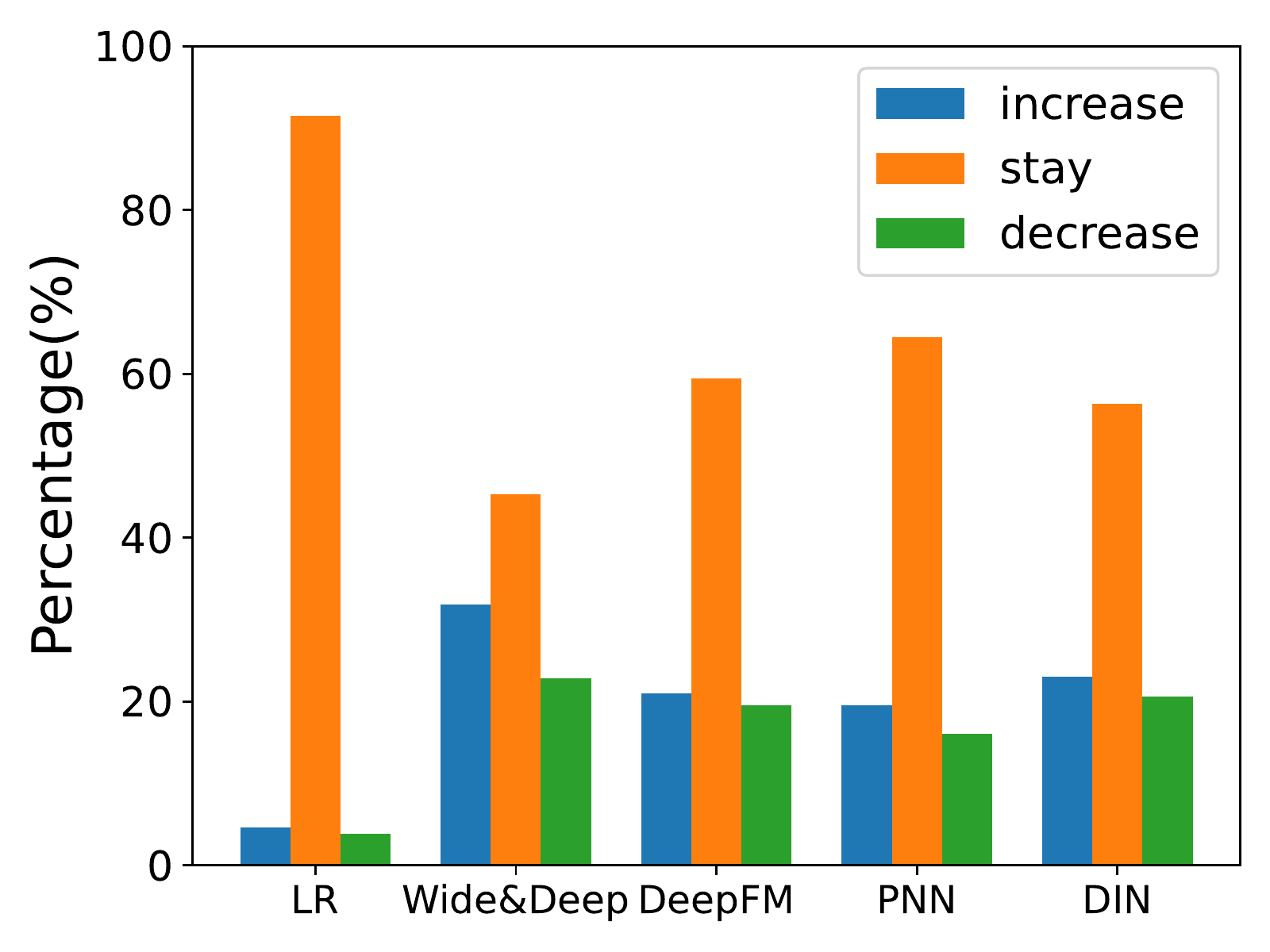}
}
\caption{Proportions of users whose test AUCs increase, stay the same, or decrease compared with cloud-based learning.}\label{lmt_stats}
\end{figure}

We finally present the evaluation results. We first show the average test AUCs of LCFT and the baselines as well as the improvement in Table \ref{public_aucs}. We can observe that LCFT outperforms all the baselines over five models and two public datasets. In contrast, purely local fine-tuning is even worse than cloud-based training on several tasks due to the CTR drift issue. These results demonstrate the necessity of label correction for effective on-device fine-tuning. We further evaluate the effect of different CTR drifts on the performance of LCFT and plot the results in Figure \ref{datasets_gaps}. The global CTRs in the MovieLens and Amazon experiments are 0.51 and 0.17, respectively. For the experiments over the MovieLens dataset, we divide the local CTRs into 10 intervals, and for each interval, we average the improvements of LCFT over purely local fine-tuning for those users whose local CTRs fall into the interval. For the experiments over the Amazon dataset, the negative samples are generated by sampling, and thus the local CTRs are discrete. We group the users based on their negative sampling ratios, and for each group, we still average the improvements of LCFT for all the users in the group. From Figure \ref{datasets_gaps}, we can observe that as the gap between the local CTR and the global CTR becomes larger (i.e., with higher CTR drift), the improvement of LCFT over the purely local fine-tuning generally becomes more evident, especially in the experiments over the Amazon dataset. Such a key observation reveals that the effectiveness of local fine-tuning indeed suffers from the CTR drift. We also depict in Figure \ref{lmt_stats} the proportions of the users whose test AUCs increase, stay the same, and decrease, thereby demonstrating the advantage of LCFT over the cloud-based learning at the user level. We can see that the proportion of the users whose AUCs increase is significantly higher than the proportion of the users whose AUCs decrease.

\subsection{Evaluation on Mobile Taobao Dataset}

\begin{figure}[!t]
     \centering
     \includegraphics[width = 0.5\columnwidth]{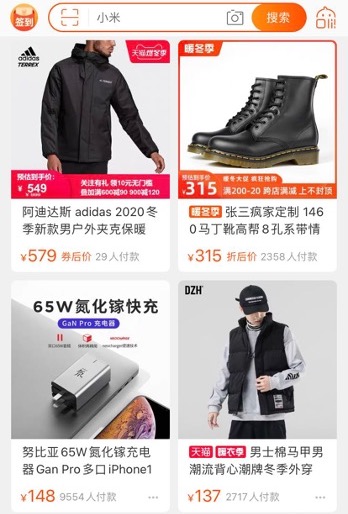}
     \vspace{0.5em}
     \caption{Homepage recommendation of Mobile Taobao.} \label{tb_app}
     \vspace{-0.5em}
\end{figure}

We also collect a practical dataset from Mobile Taobao and conduct offline experiments. In what follows, we introduce the application scenario, the dataset, the experimental setups, and the evaluation results.

The application scenario is the homepage of Mobile Taobao, where different items are ranked based on each user's preferences, as shown in Figure \ref{tb_app}. The rule of ranking the items depends on the CTRs of them. Therefore, the basic learning task is CTR prediction. The input data fields include user feature, item exposure user action feature, item page-view user action feature, and item feature. We take the recommendation model, called EdgeRec~\cite{net_edgerec}, which has been offloaded to the Mobile Taobao APP for on-device inference. EdgeRec contains an embedding layer, a GRU layer, an attention layer, and fully connected layers (i.e., MLP layers). In particular, the GRU layer first encodes the user exposure and behavior sequences, and then the target item is on attention with the encoded sequences. 


The dataset is collected from March 1, 2021 to March 7, 2021, in total 7 days. For the split of each user’s training and test sets, we take the first 5 days of samples for training and take the last 2 days of samples for testing. We keep the pool of 30,682 users who have more than 256 training samples. 

Regarding the fine-tuning settings, we use Adam as the optimization algorithm. We set the batch size to 32 and let each user train 3 epochs with the learning rate of 0.001. 

We finally show the user-level average test AUCs in the last row of Table 1. We can observe that the evaluation results of the Mobile Taobao dataset are consistent with those over the public datasets, namely, LCFT outperforms all the baselines. In contrast, the average test AUC of local training decreases a little compared with the cloud-based learning model, which demonstrates the necessity of label correction in industrial application scenarios.

\subsection{Online A/B Testing in Mobile Taobao}

We deploy LCFT on different groups of Mobile Taobao users and report the online A/B testing results from August 1, 2021 to August 6, 2021.

For online A/B testing, we create two non-overlapping buckets, each of which contains roughly 150,000 randomly chosen users, to deploy LCFT and the cloud-based learning model. In particular, the cloud-based model is trained over a 7-day dataset with billions of samples. For the deployment of LCFT, we cluster the users into three groups based on their local CTRs (i.e., the users with low, middle, and high local CTRs). For each user group, the cloud-based global model is fine-tuned using LCFT over the training data from the users in the group and then offloaded to the corresponding mobile devices for real-time inference. 

We use three online metrics to extensively evaluate the performance of LCFT and the cloud-based learning. Before giving formal definitions, we first introduce some common abbreviations in recommender systems: ``Clk'' is short for ``click''; ``Exp'' is short for ``exposure''; ``UV'' is short for “Unique Visitor”. Then, the three metrics are defined as follows: (1) {\text{Clk}}/{\text{Exp}} is in fact CTR and is the optimization objective of the learning task in our application scenario; (2) and {\text{Clk}}/{\text{UV}} denotes the average number of clicks per user, which can measure the user activeness.

\begin{table}[!t]
    \caption{Online A/B testing results in Mobile Taobao. Metrics are reported in the day-level average.}
    \label{abtest}
    \centering
    \resizebox{0.9\columnwidth}{!}{
    \begin{tabular}{cccc}
    \toprule
        Metric & Cloud & LCFT & LCFT vs. Cloud \\
        \midrule
        CTR & 3.91\% & 3.98\% & {\bf +1.79\%}  \\
        \midrule
        Clk/{\text{UV}} & 4.60 & 4.64 & {\bf + 1.03\%} \\
    \bottomrule
    \end{tabular}
    }
\end{table}

\begin{figure}[!t]
        \centering
        \subfigure[CTR]{
        \includegraphics[width=0.472\columnwidth]{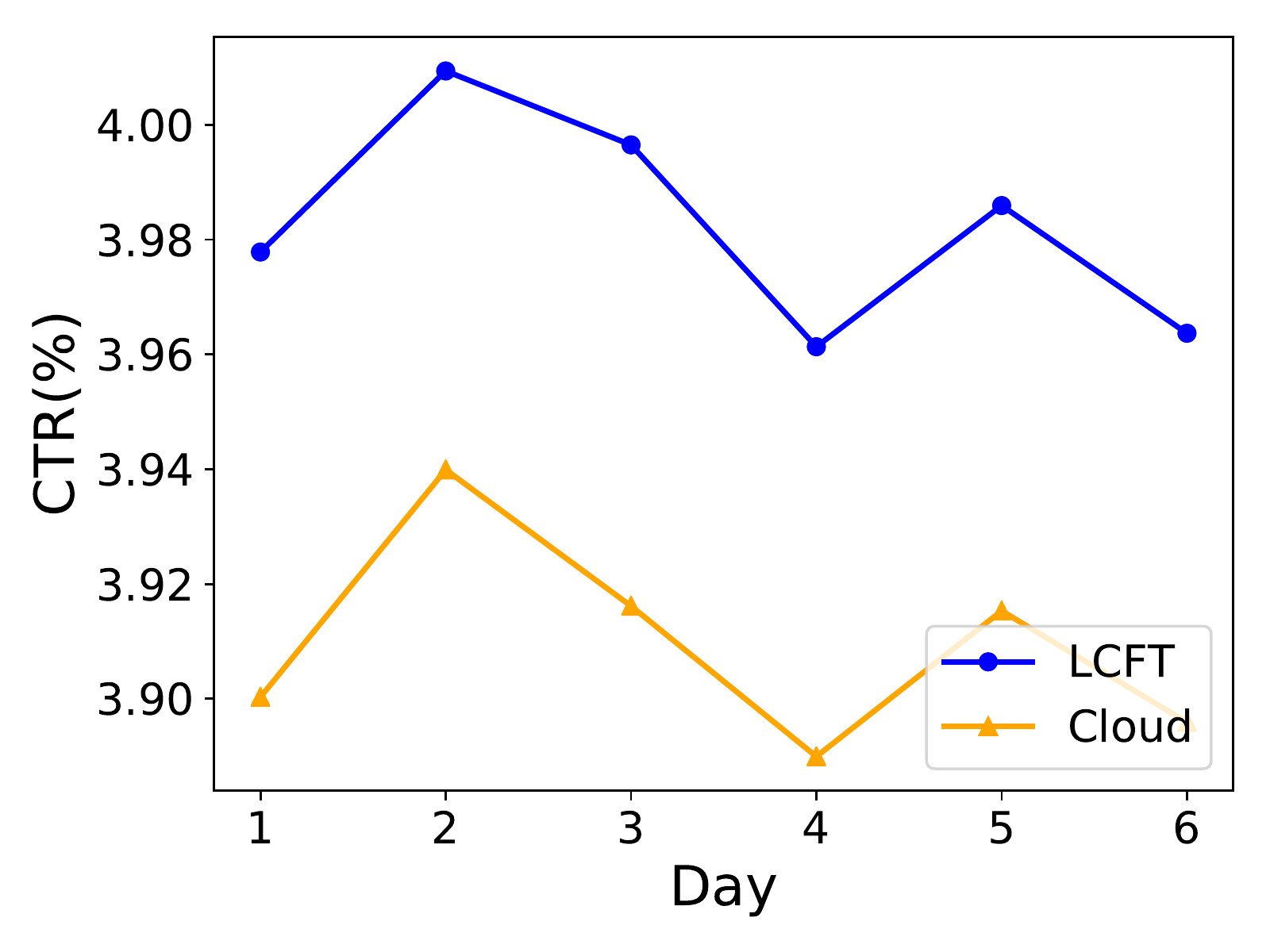}
        }
        \subfigure[Clk/UV]{
        \includegraphics[width=0.472\columnwidth]{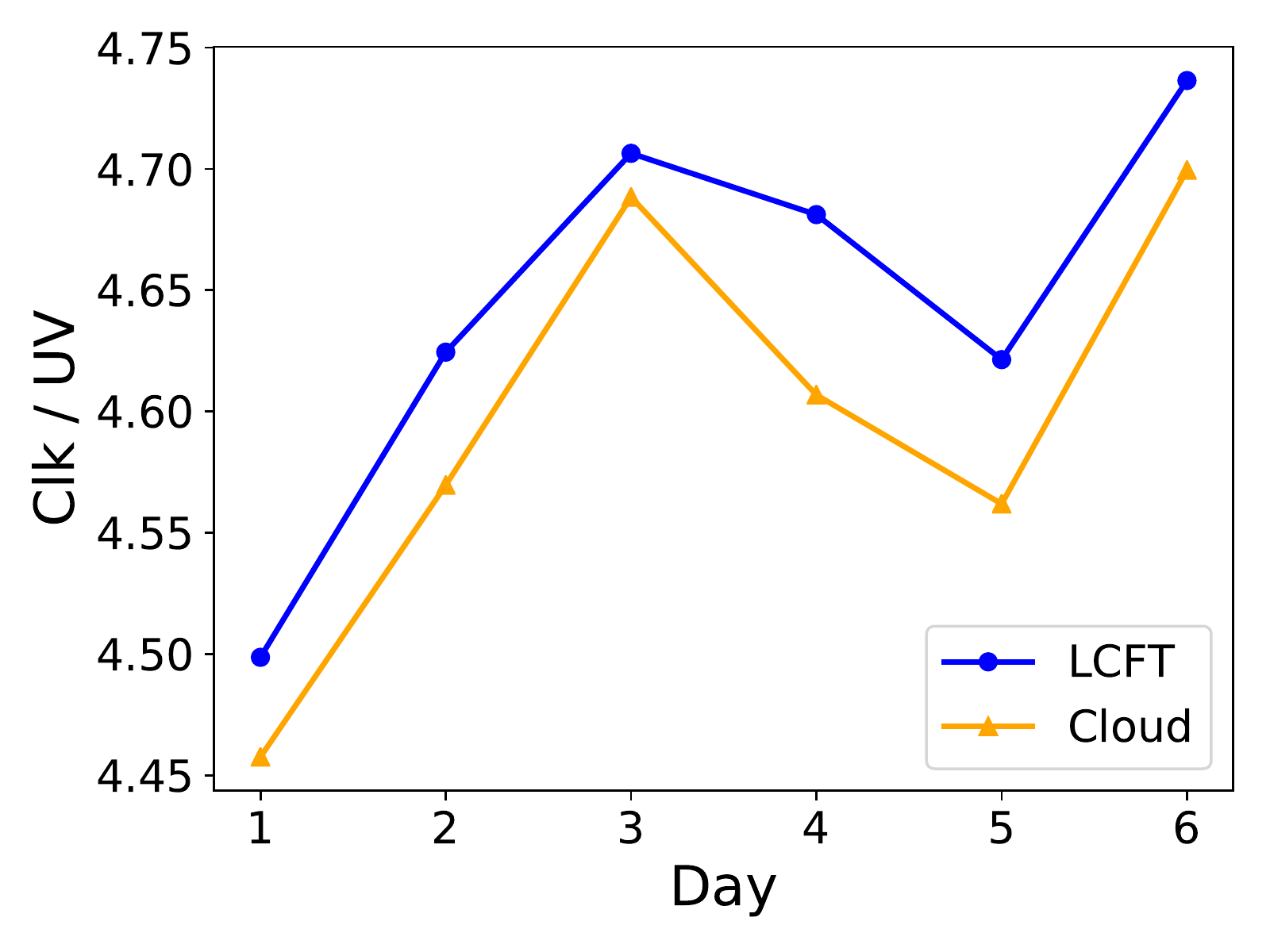}
        }
        \caption{Online performance of LCFT and cloud-based learnaing in Mobile Taobao.}
        \label{online_fig}
\end{figure}

We finally plot in Figure \ref{online_fig} the day-level results from the online metrics and summarize in Table \ref{abtest} the day-level average results. Compared with cloud-based learning, LCFT improves CTR, {\text{Clk}}/{\text{UV}} by 1.79\% and 1.03\%, respectively. These results demonstrate that through fine-tuning with label correction, LCFT indeed improves the online recommendation performance in practice.

\section{Conclusion}
In this work, we have proposed an on-device model fine-tuning with label correction method, called LCFT, for the fundamental CTR prediction task in recommender systems. LCFT mitigates the bottleneck issue of CTR drift through only letting each user correct the labels of the local samples ahead of fine-tuning, thereby aligning the equivalent CTR with the global CTR in theory. Offline and online evaluation results have validated the necessity of label correction and demonstrated the superiority of LCFT over the mainstream cloud-based learning method without fine-tuning.

\bibliography{lcft}

\end{document}